\documentclass[12pt]{article}
\usepackage{amsfonts,epsfig}
\usepackage{amsmath,amsfonts}
\usepackage{amssymb,graphicx}

\def\pa{\partial}

\def\p{\varphi}

\def\h{\hat}
\def\b{\beta}

\def\f{\frac}

\def\p{\varphi}

\def\h{\hat}

\def\f{\frac}

\def\l{\label}

\def\a{\alpha}

\def\g{\gamma}
\def\G{\Gamma}
\def\m{\mu}
\def\n{\nu}

\def\r{\rho}

\def\S{\Sigma}

\def\th{\theta}
\def\o{\omega}
\def\O{\Omega}

\def\be{\begin{equation}}
\def\ee{\end{equation}}
\def\ba{\begin{eqnarray}}
\def\ea{\end{eqnarray}}

\begin{document}

\begin{center}
{\large \bf Thermal Hawking radiation of black hole with supertranslation field}
 \end{center}
\vspace*{1cm}
\centerline{\bf Mikhail Z. Iofa
\footnote{ e-mail:iofa@theory.sinp.msu.ru}}
\centerline{Skobeltsyn Institute of Nuclear Physics}
\centerline{Moscow State University}
\centerline{Moscow 119991, Russia}

\begin{abstract}

Using the analytical solution for the Schwarzschild metric
containing supertranslation field, we consider  
 two main ingredients of calculation of the thermal Hawking black hole radiation:
solution for eigenmodes of the d'Alambertian and solution of the geodesic 
equations for null geodesics.
For calculation of Hawking radiation it is essential to determine the
behavior of both the eigenmodes and geodesics in the vicinity of
horizon. The equation for the eigenmodes is solved, first,
perturbatively in the ratio $O(C)/M$ of the supertranslation field
to the mass of black hole, and, next, non-perturbatively in the near-
horizon region. 
It is shown that in any order of perturbation
theory solution for the eigenmodes in the metric containing 
supertranslation field differs
from solution in the pure Schwarzschild metric by terms of order
$L^{1/2}= (1-2M/r)^{1/2}$. 
In the non-perturbative approach, solution for the eigenmodes 
differs from solution in the Schwarzschild metric by terms of order
$L^{1/2}$ which vanish on horizon. Using the simplified form of 
geodesic equations  in 
vicinity of horizon, it is shown that in vicinity 
of horizon the null geodesics  have the same behavior 
as in the Schwarzschild metric. As a result,
the density matrices of thermal radiation in both cases are the same.  

\end{abstract} 	

\section{Introduction}

Recently there was a renewed interest in asymptotic symmetries at the
null infinity, the BMS symmetries \cite{bondy,sachs}.
The group of symmetries of asymptotically flat gravity, the BMS group,
 extends the Poincare
group and is the semi-direct product of the Lorentz group and the normal
abelian subgroup of supertranslations which generalize translations.
The $BMS$ group
can be enlarged  to a group which also contains superrotations,
singular supertranslations and local conformal transformations
\cite{bar1,bar3,bar2}.

When acting on an asymptotically flat physical state
the $BMS$ group
of diffeomorphisms transforms it to another physical state
preserving the asymptotic flatness.
The BMS group contains coordinate
transformations which are pure gauge transformations and also diffeomorphisms
that change supertranslation field in the metric and provide the
 mapping of a physical state to another physical state \cite{strom1,comp2}.

In paper \cite{strom1}, was obtained an important  physical
result  that provided a
metric decreases fast enough at infinity, in some neighborhood of Minkowski
vacuum the $S$ matrix is invariant under an infinite-dimensional
 subgroup of $BMS^+ \times BMS^-$,
where $BMS^\pm$ are the groups acting at the future and past null infinities.
The correlated result 
is that in the low energy gravitational scattering, 
because of the supertranslation invariance of the S-matrix,
the local energy is conserved
at any angle. In papers  \cite{strom1,strom2} connection of an 
(in principle) observable gravitational memory effect 
with the change of vacua under the action
of the BMS transformations was elucidated and connection with the soft graviton
theorem \cite{winb} was established. Possible relation of  the extended
BMS group to the black hole information problem was discussed in \cite{hawk}.

In papers \cite{comp1,comp2} a method to construct a metric 
containing a finite supertranslation field in the bulk was
developed. The metric was obtained by exponentiation of infinitesimal
supertranslation diffeomorphism. When applied to the Schwarzschild metric,
solution generating technic of \cite{comp2,comp1} provides generalization 
of the Schwarzschild metric containing  supertranslation field $C(\th ,\p )$.
 
Using the analytical solution of paper \cite{comp2}
 we consider calculation of "hard" thermal Hawking radiation \cite{strom3} 
of the Schwarzschild black hole with supertranslation field. 
In the present paper we consider the
 supertranslation field depending only on $\th :\,\, C=C(\th )$.
Calculations are performed perturbatively in the ratio
$O(C)/M$, where $M$ is the mass of
black hole. 

Following the standard procedure of calculation of the Hawking radiation
which assumes that the radiated particles are produced in the near-horizon 
region \cite{haw,bir,brout}, we study  
solutions for the eigenmodes of the d'Alambertian in the limit $r\rightarrow 2M$. 
We find that in any order in perturbation theory in $O(C)/M$, 
in the vicinity of horizon, solutions for 
the eigenmodes in the metric with supertranslation field differ from solutions
in the pure Schwarzschild background by the terms of order
 $L^{1/2}O(C^n /M^n )$ where $L=(1-2M/r)$. 
Next, we show that the same result is valid 
for the non-perturbative solution, in which case the difference from 
 the zero-order
solution $\psi_0$ is $L^{1/2}\p (\th ,r )\psi_0$, where $L^{1/2}\pa_r \p$ is
finite in the limit $L=0$.

Evolution of massless field is determined by its data on the past null infinity.
To calculate the particle content at the future infinity in terms of 
excitations at the past infinity,
the outgoing modes are traced back to the past null infinity and
expanded in the basis of the incoming modes.
To trace the outgoing modes to the past null infinity, we
consider  the  null geodesics depending on $r$ and $\th$ 
in the Schwarzschild background 
with the supertranslation field .
The geodesic equations simplify in the vicinity of the horizon.
We find that in the near-horizon region behavior of radial null geodesics 
 is similar to those in the Schwarzschild background.

Collecting these results we find that the Bogolubov coefficients 
and the density matrix of thermal radiation in the metric with supertranslation
field are the same as in the pure Schwarzschild background.

\section{Generalization of the Schwarzschild metric containing the
supertranslation field}

The generic final state of collapse of matter is a stationary space-time
which metric is diffeomorphic to the Kerr metric, if deviation from the Kerr
metric is small.
In \cite{comp2} was constructed a metric generalizing 
the Schwarzschild metric and
containing  supertranslation
field. In the isotropic spherical coordinates  in which the original
Schwarzschild metric is written as
\be
\l{1.1}
ds^2 =-\f{(1-M/2\r )^2}{(1+M/2\r )^2}dt^2 +
(1+M/2\r )^4 (d\r^2 +\r^2 d\O^2 ),
\ee
the  metric containing the supertranslation
field  $C(\th ,\p )$ was obtained in a form
\be
\l{2.1}
ds^2 =-\f{(1-M/2\r_s )^2}{(1+M/2\r_s )^2}dt^2 +
(1+M/2\r_s )^4 \left(d\r^2 + (((\r -E)^2 +U)\g_{AB} +(\r -E)C_{AB})dz^A dz^B
 \right).
\ee
Here
\be
\l{4.1}
\r_s (\r, C) =\sqrt{(\r -C -C_{00})^2 + D_A C D^A C}
.\ee
$C_{00}$ is the lowest constant spherical harmonic of $C(\th ,\p )$.
In the following we do not write $C_{00}$ explicitly understanding
$C\rightarrow C-C_{00}$.
Covariant derivatives $D_A $ are  defined
with respect to the metric on the sphere
 $ds^2 =d\th^2 +\sin^2\th d\p^2$.
For technical simplicity we consider the case of $C$ depending only on
$\th ,\,\,\, C=C(\th )$.

The tensor $C_{AB}$  and the functions
$U$ and $E$ are defined as
\ba
\l{3.1}
\nonumber
C_{AB} = -(2D_A D_B -\g_{AB} D^2 ) C,\\
U=\f{1}{8} C_{AB}C^{AB},\\\nonumber
E=\f{1}{2}D^2 C + C
.\ea
The function $C$ has dimension of mass.

The non-zero Christoffel symbols  are $\G^\p_{\p\th}=\cos\th/\sin\th,\,\,\,
\G^\th_{\p\p}=-{\cos\th}{\sin\th}$. For $C_{AB}$ we obtain
\ba
&\l{6.1}&C_{\th\th}=-\left(C'' -C'\cot\th \right)\\
&\l{7.1}&C_{\p\p} =\sin^2\th\left(C''-C'\cot\th\right)\\\nonumber
&{}&C_{\p\th}=0
\ea
Here prime denotes the derivative $\pa_\th$.
From (\ref{6.1})-(\ref{7.1}) it follows that $C_{\p\p}=-\sin^2\th C_{\th\th}$.
The functions $E(\th)$ and $U$ are
\ba
\l{9.1}
&{}&E=\f{1}{2}\left(C''+C' \cot\th\right)+C,\\\nonumber
&{}&U=\f{1}{4}\left(C''-C' \cot\th\right)^2.
\ea

We assume that $C(\th )$ is such that the components of the metric 
are finite.  In particular, this condition is fulfilled for 
$C(\th )=\sum a_n\,P_n (\cos \th)$, where $P_n (x)$ are 
Legendre polynomials.
Note that despite the term $\cot\th$ contained in the
functions (\ref{6.1}), (\ref{7.1}), the components of the metric 
(\ref{e.4}), (\ref{e.5}) are finite.

Equation $\r_s (\r_H ,C) = M/2$ defines  location of the 
horizon $\r_H$. For  horizon to exist, condition
$M^2 /4 - D_A C D^A C >0$ should be fulfilled.
In the case $C = C(\th )$ condition of existence of
horizon takes the form $M/2 > | C'(\th )|$.

Determinant of the angular part of the metric (\ref{2.1}), $(\r-E)^2 -U$,
vanishes on the surfaces \cite{comp2}
$$\r_{SH\pm} =E\pm\sqrt{U}= \f{1}{2}(\left(C''+C' \cot\th\right) 
\pm\f{1}{2}\left(C''-C' \cot\th\right) 
.$$
The surfaces $\r_{SH\pm} $ are located in the
 region  $\r < \r_H$.
  
The class of models with a supertranslation fields $C(\th )$ contains the 
Kerr solution in which case $C_{\th\th}(\th )= a/\sin\th$ \cite{flet,bar4}
and $C(\th )=a_2 +a_1\cos\th +a\sin\th$ with $a_1 , a_2$ arbitrary.

Let us transform the metric (\ref{2.1}) to the Schwarzschild variables
introducing
\be
\l{11.1}
r=\r_s (\r, C)\left(1+\f{M}{2\r_s (\r, C)}\right)^2
.\ee
Inverting this equation, we obtain
\be
\l{12.1}
\r_s (\r, C) = \f{1}{2}(r-M+\sqrt{r(r-2M)})
\ee
In variable $r$ horizon is located at the point $r =2M$.

Let us introduce  notations
\ba
\l{13.1}
&{}&L=1-\f{2M}{r},\\
\l{14.1}
&{}&K=r-M + r L^{1/2}, \qquad \,\,\,\f{dK}{dr}= \f{K}{rL^{1/2}}.
\ea
Using the relation (\ref{12.1}), we have
\be
\l{13.11}
\f{(1-M/2\r_s )^2}{(1+M/2\r_s )^2}=L,\qquad\,\,
(1+M/2\r_s )^4 =\f{4r^2}{K^2}.
\ee
 From the equation (\ref{12.1}) written as
$$
\sqrt{(\r -C)^2 + D_A C D^A C} =\f{K}{2},
$$
 we obtain the expression for $\r$ as a function of $r$ and $C$
\be
\l{12.11}
\r = C +\f{K}{2}\left(1-\f{4(DC)^2}{K^2}\right)^{1/2}.
\ee
Let us denote
\be
\l{e.1}
b=\f{2C'}{K}
,\ee
so that (\ref{12.11}) becomes
\be
\l{e.7}
\r =C+\f{K}{2}\sqrt{1-b^2}
.\ee
Because $K$ is the increasing function of $r, \,\, b $ has its maximum at
$r=2M$ equal to $2|C'|/M$. Relation (\ref{e.7}) is meaningful for $|C'|< M/2$.

Differentiating (\ref{12.11}), we obtain 
\be
\l{e.2}
d\r =\f{K}{2}\left[\left(b-\f{bb'}{\sqrt{1-b^2}}\right)d\th+
\f{dr}{rL^{1/2}\sqrt{1-b^2}}\right].
\ee
Let us consider the components $g_{\th\th}$ and $g_{\p\p}$ 
in the metric (\ref{2.1}).
 Noting that
$U =C_{\th\th}^2 /4$, we obtain
\ba
\l{e.4}
g_{\th\th}=((\r -E)^2 +U)\g_{\th\th} +(\r -E)C_{\th\th}=
\left(\r-E+\f{C_{\th\th}}{2}
\right)^2=
\f{K^2}{4}(\sqrt{1-b^2}-b')^2 {}{}{}{}    \\
\l{e.5}
g_{\p\p}=((\r -E)^2 +U)\g_{\p\p} +(\r -E)C_{\p\p}
=\left(\r-E-\f{C_{\th\th}}{2}
\right)^2\sin^2\th =
\f{K^2}{4}\sin^2\th(b\cot\th -\sqrt{1-b^2})^2.
\ea
The surfaces at which $g_{\th\th}$ and $g_{\p\p}$ vanish 
are located in the intrnal domain $r>2M$.

Using the expressions (\ref{13.1}), (\ref{13.11}) and  substituting the 
components of the metric
 (\ref{e.4}), (\ref{e.5}) in (\ref{2.1}),
we obtain the metric in Schwarzschild variables 
\ba
\l{e.6}
ds^2 = -Ldt^2 + \f{dr^2}{L(1-b^2 )} +2dr d\th \f{b(\sqrt{1-b^2}-b')r}
{(1-b^2 )L^{1/2}} \\\nonumber
+ d\th^2 r^2 \f{ (\sqrt{1-b^2}-b')^2}{1-b^2} +
d\p^2 r^2\sin^2\th (b\cot\th -\sqrt{1-b^2})^2 .
\ea
Determinant of the metric (\ref{e.6}) is
\be
\l{e.8}
|g|= r^4\sin^2\th\f{ (\sqrt{1-b^2} -b')^2 }{1-b^2} 
 (\sqrt{1-b^2}-b\cot\th )^2
.\ee
Because of conditions $M/2 >(|C'|, \r_{SH} )$ we have $|g|>0$ for all $\th$ and
$r>2M$.
The inverse metric is
\ba
\l{e.9}
\left(\begin{array}{cccc}
-L^{-1}&0&0&0\\[2mm]
0&L&-L^{1/2} b[r(\sqrt{1-b^2}-b')]^{-1}&0\\[2mm]
0&-L^{1/2} b[r(\sqrt{1-b^2}-b')]^{-1}&[r (\sqrt{1-b^2}-b') ]^{-2}&0\\[2mm]
0&0&0& [r^2\sin^2\th(\sqrt{1-b^2} -b\cot\th \,)^2]^{-1}
\end{array}\right)
\ea
 
We introduce the following notations which we use below
\ba
\l{e.10}
&{}&\sqrt{|g|} = \sqrt{g_0 |\h{g}|}=  r^2 \sin\th\sqrt{|\h{g}|},\\\nonumber 
&{}&\sqrt{|\h{g}|}= 1+g_1 +g_2 +\cdots =\\
&{}&  1- (b' +b\cot\th ) - \left(\f{b^2}{2} -b'b \cot\th\right )+\cdots
= 1-\f{D^2 C}{K} +\f{(D^2 C )^2 -C_{\th\th}^2 -2(DC)^2}{K^2} +\cdots
\\
\l{e.11}
&{}& g_{rr}= \f{1}{L}\h{g}_{rr},\qquad \,\, 
g_{r\th}=\f{L^{1/2}}{r}\h{g}_{r\th},\qquad\,\,
g^{r\th}=-\f{r}{L^{1/2}}\h{g}^{r\th},\qquad g_{\th\th}=r^2\h{g}_{\th\th}, \\
\l{e.12}
&{}&\h{g}^{r\th} = g^{r\th}_1 +g^{r\th}_2 +\cdots =
b +bb'+\cdots=
\f{2C'}{K} + \f{2{(DC)^2}'}{K^2}+\cdots,\\
\l{e.13}
&{}&f_1 =\sqrt{1-b^2}-b',\\
\l{e.14}
&{}&f_2 =\sin^2\th (\sqrt{1-b^2}-b\cot\th ).
\ea

\section{Equation for eigenmodes}

In this section  we study solutions for the eigenmodes of the equation 
$\Box (g) \psi =0$ perturbatively and non-perturbatively
 in small parameter $O(C)/K \sim O(C)/M$
and prove that corrections to solutions in the pure Schwarzschild metric
 vanish in the vicinity 
of horizon $r=2M$ as $L^{1/2}= (1-2M/r )^{1/2}$.

The operator $\Box (g)$ is
\ba
&{}&\Box  =\f{1}{\sqrt{-g}}\pa_\m\sqrt{|g|}g^{\m\n}\pa_\n =\\\nonumber 
&{}&g^{tt}\pa_t^2 +\f{1}{\sqrt{-g}}\pa_r \sqrt{-g} g^{rr}\pa_r
+\f{1}{\sqrt{-g}}\pa_r \sqrt{-g} g^{r\th}\pa_\th +
\f{1}{\sqrt{-g}}\pa_\th \sqrt{-g} g^{\th r}\pa_r  +\\\nonumber
&{}+&\f{1}{\sqrt{-g}}\pa_\th \sqrt{-g} g^{\th\th}\pa_\th+
\f{1}{\sqrt{-g}}\pa_\p \sqrt{-g} g^{\p\p}\pa_\p,
\ea
or explicitly
\ba
\l{d.1}
&{}&\Box\psi = 
-L^{-1}\pa^2_t \psi+\f{1}{\sqrt{-g}}\pa_r \sqrt{-g} L\pa_r\psi
- \f{1}{\sqrt{-g}}\pa_\th 
  \sqrt{-g}\f{L^{1/2}b}{r\,f_1}\pa_r\psi\\\nonumber
&-& \f{1}{\sqrt{-g}}\pa_r \sqrt{-g}
  \f{L^{1/2}b}{r\,f_1}\pa_\th\psi
+ \f{1}{\sqrt{-g}}\pa_\th \sqrt{-g}  (r\, f_1 )^{-2} 
\pa_\th\psi +
\f{1}{\sqrt{-g}}\pa_\p \sqrt{-g} 
(r\, f_2 )^{-2} \pa_\p\psi=0
.\ea
First, we solve the equation $\Box\psi =0$ perturbatively in $O(C)/K$  
taking $\psi =\psi_0 +
\psi_1 +\psi_2+\cdots$ and $\Box =\Box_0 +\Box_1 +\Box_2 +\cdots$,
where subscripts  denote the  orders in $O(C)/K$. We have
\be
\l{d.2}
 \Box \psi = \Box_0 \psi_0 +(\Box_1 \psi_0 +\Box_0 \psi_1 ) +(\Box_2 \psi_0 
+\Box_0 \psi_2 +\Box_1 \psi_1 ) +\cdots. 
\ee

Expanding the terms in (\ref{d.1}) in the series in  $O(C)/K$, and using notations 
(\ref{e.10})- (\ref{e.14}), we have
\ba 
\l{nd.1}
&{}&\f{1}
{\sqrt{-g}}\pa_r \sqrt{-g} L\pa_r\psi = \left[L\pa_r +
L(2/r) + (\pa_r L) +L(\pa_r g_1 )  +
L((\pa_r g_2 -g_1 (\pa_r g_1 ) )+\cdots \right] \pa_r\psi
\\\nonumber
&{}& 
\f{1}{\sqrt{-g}}\pa_\th \sqrt{-g}
  \f{L^{1/2}b}{r\,f_1}\pa_r\psi =
\left(\f{L^{1/2}}{r}\right)\left[
g_1^{r\th}\pa_\th +(\cot\th g_1^{r\th} +(\pa_\th g_1^{r\th}) ) +
\right.
\\\l{nd.2}
&{}&\left.(g_2^{r\th}\pa_\th +\cot\th g_2^{r\th} +(\pa_\th g_2^{r\th})  
 +(\pa_\th g_1 )g_1^{r\th})+\cdots\right]\pa_r\psi
\\\nonumber
&{}&
\f{1}{\sqrt{-g}}\pa_r \sqrt{-g}
 \f{L^{1/2}b}{r\,f_1}\pa_\th\psi =\left(\f{L^{1/2}}{r}\right)
\left[g_1^{r\th}\pa_r +(2/r)g_1^{r\th} +(\pa_r g_1^{r\th})\right.+
\\\l{nd.3}
&{}&\left.(g_2^{r\th}\pa_r +(2/r)g_2^{r\th} +(\pa_r g_2^{r\th}) +
 (\pa_r g_1 )g_1^{r\th} ) +\cdots\right]\pa_\th\psi
\ea

Numerical calculations \cite{sanch}
of the Hawking radiation in different modes and theoretical considerations
\cite{viss} have shown that the main part of the energy 
of radiation is contained in the $s$-wave. 
Because of that, first, in the main order, we consider the $s$-
wave eigenmode and, next, at the end of the Section, comment on
 the modes with $l>0$.

In the zero order in $O(C)/K$, written  
in tortoise variable $r_* =r +2M \ln (r/2M -1)$, the operator $\Box_0 \psi$ is
\be
\l{d.3}
\Box_0 \psi =(rL)^{-1}\left[-\pa_t^2 +\pa_{r_*}^2 +L\left (\f{2M}{r^3} -
\f{\h{K}^2 (\th,\p )}{r^2}\right)\right]r\psi,
\ee
where $\h{K}^2$ is the angular momentum operator.
In calculation of the Hawking radiation it is important to determine the
behavior of the eigenmodes in the vicinity of the horizon \cite{haw,brout,viss}.
We look for solution of the equation for the eigenmodes
in the region  $r-2M \ll M$ and set $r=2M$ in the
functions with regular behavior at $r=2M$.

When acting on $\psi_0$ the derivatives $\pa_r$ produce the terms with 
powers of $L^{-1}$ which grow up  as $r\rightarrow 2M$. 
In the leading order in $L^{-1}$ we have
\be
\l{d.7}
\pa_r \f{e^{i\o r_*}}{r}\simeq \f{e^{i\o r_*}}{r}
\f{i\o}{L},\qquad
\pa_r^2 \f{e^{i\o r_*}}{r}\simeq -\left(\f{\o^2}{L^2} +
\f{2iM\o}{L^2\, r^2}\right)
\f{e^{i\o r_*}}{r}.
\ee
Acting on $O(C^n )/K^n$ derivatives $\pa_r$ produce the factor $L^{-1/2}$
\be
\l{d.71}
\pa_r \f{O(C^n )}{K^n}= -n\f{O(C^n )}{K^n rL^{1/2}}
.\ee

In the region $r\sim 2M$, neglecting in the operator $\Box_0$
the small term $L2M/r^3$, solution
of the Eq. (\ref{d.3}) for the	$s$-mode is obtained as
\be
\l{d.4}
\psi_0 = \f{e^{i\o (t \pm r_* )}}{\sqrt{4\pi} r}.
\ee

Let us consider the equation (\ref{d.2}) in the first order
 $\Box_1 \psi_0 + \Box_0 \psi_1=0$.
Using (\ref{nd.1}),  (\ref{nd.2}) and (\ref{e.10}), (\ref{e.11}), 
we obtain (\ref{k.2}) 
 the terms without the derivatives 
$\pa_\th\psi_0$ and $\pa_\p\psi_0$ in the operator $\Box_1\psi_0$ as  
\ba
\l{d.5}\nonumber
&{}&\left[L(\pa_r g_1)+\left(-\f{L^{1/2}}{r}\right)(\cot\th g_1^{r\th} +
(\pa_\th g_1^{r\th})\right]\pa_r\psi_0=
\\
&{}&=
\left[L\pa_r(-b' -b\cot\th )- 
\left(\f{L^{1/2}}{r}\right)(b' +b\cot\th )\right]\pa_r\psi_0=0
\ea
where we used the formula (\ref{d.71}) to obtain
$L\pa_r b =L\pa_K (2C/K)\pa_r K =-bL^{1/2}/r$.
The remaining terms in the operator $\Box_1 \psi_0 $ contain derivatives $\pa_\th$
and $\pa_\p$
acting on $\psi_0$ and vanish. The equation $\Box_1 \psi_0 +\Box_0 \psi_1=0$
 reduces to  $\Box_0 \psi_1 =0$ and yields $\psi_1 =\psi_0$.

In the second order  in $O(C )/K$  the equation (\ref{d.2}) is
$\Box_2\psi_0 +\Box_1\psi_1 + \Box_0\psi_2 =0$ which reduces to 
$\Box_2\psi_0 + \Box_0\psi_2 =0$.
 The terms without derivatives over angular variables
acting on $\psi_0$ are
\be
\l{d.6}
\Box_2\psi_0=
\left[L((\pa_r g_2 -g_1 (\pa_r g_1 ) )+
\left(-\f{L^{1/2}}{r}\right)
(\cot\th g_2^{r\th}+(\pa_\th g_1 )g_1^{r\th} 
+(\pa_\th g_2^{r\th})\right]\pa_r\psi_0
\ee
Both terms in the square brackets are of order $L^{1/2}$.
In the leading order in $L^{-1}$ we have
\be
\l{d.8}
\Box_2\psi_0 = \f{\pm i\o }{ L^{1/2}}\f{F_{(2)}(\th , r )}{K^2}\psi_0,
\ee(\ref{k.2})
where $F_{(2)}=O(C^2)$, and the explicit form of $F_{(2)}$ is irrelevant for us.
 The equation $\Box_0 \psi_2 = -\Box_2 \psi_0$
 has the following structure
\be
\l{d.9}
\f{1}{rL}\left[-\pa_t^2 +\pa_{r_*}^2 +
L\left(\f{2M}{r^3} -
\f{\h{K}^2 (\th,\p )}{r^2}\right)\right]
r\psi_2
= \f{\mp i\o}{L^{1/2}} \f{F_{(2)}}{K^2} \f{e^{i\o (t\pm r_* )}}{r}
\ee
Looking for a solution in the form $r\psi_2 = f e^{i\o (t\pm {r_*})}$,
in the leading order in $L^{-1}$
 we obtain
\be
\l{rd.3}
\pa_{r_*}^2 f \pm 2i\o \pa_{r_*}f = \pm i\o L^{1/2}\f{F_{(2)}}{K^2}
\ee 
In the region $r\simeq 2M$ approximately $r_* =2M\ln L$. Solving the
equation (\ref{rd.3}), we have
\be
\l{rd.4}
\psi_2= \pm i\o\f{F_{(2)}}{K^2}\f{L^{1/2}(4M)^2}{1\pm 8i\o M}\psi_0
.\ee

In the higher orders we proceed by induction. 
The equation for $\psi_n$ is
\be
\l{rd.5}
\Box_0 \psi_n +\sum_{k+r=n,\, k,r \neq 0}  \Box_k \psi_r + \Box_n \psi_0=0
.\ee
Let $\psi_r$ have the following structure
\be
\l{rd.6}
\psi_r =L^{1/2}F(O(C^k /K^k) ,\th, r)\psi_0,\qquad\,\,k\lesssim r,
\ee
The operators $\Box_k ,\,\, k>2$ can be presented in a form
\be
\l{rd.7}
L(\pa_r F_1 ) \pa_r + \pa_\th L^{1/2} F_2 \pa_r + \pa_r L^{1/2} F_3  \pa_\th
,\ee
where $F_i = F_i(O(C^l /K^l) ,\th, r)\sim O(L^0 )$.
When acting on the functions (\ref{rd.6}) each term in  (\ref{rd.5})
yields an expression of the form (\ref{rd.6}) without the prefactor $L^{1/2}$
i.e. the result is of order $O(L^0 )$. 
The operator $\Box_n$ is of the same structure as (\ref{rd.7}), but when acting on
the function $\psi_0$, it produces an expression of order $L^{-1/2} $.
Eq. (\ref{rd.5}) reduces to $\Box_0 \psi_n + \Box_n \psi_0 =0$ which 
is of the  functional form similar to (\ref{d.9}) and yields solution
$$
\psi_n \sim L^{1/2}O(C^n/K^n)\psi_0
.$$
The origin of this result can be traced back
to the form of d'Alambertian (\ref{d.1}) in which the $(t,r)$ part 
is the same as in the Schwarzschild metric and the additional $(r,\th)$
terms contain the factor $L^{1/2}$.

Let us consider solutions with higher harmonics, $l>0$.
 Approximate solution of the equation $\Box_0 \psi_{0l} =0$ in the 
main order in $L^{-1}$ with the $l$-th harmonic is 
$$
\psi_{l,0}=\f{e^{i\o (t \pm r_* )}}{\sqrt{4\pi} r}P_l (\cos\th )
.$$
Now the terms containing $\pa_\th\psi_{l,k}$ are nonbibitem-zero. 
In the next order, in  
the operator $\Box_1\psi_{l-1,0}$, there appears the new term
 $$
\f{1}{\sqrt{g}}\pa_r \sqrt{g}\big|_{(0)} g^{\th r}_1\pa_\th\psi_{l-1,0}=
 \f{1}{r^2}\pa_r r^2 \f{ L^{1/2}}{r}\h{g}^{\th r}_1 \pa_\th \psi_{l-1,0}
.$$
 Because $\pa_\th$ does
not change the order in $L$, this term is of order
$L^{1/2}$ and can be neglected. In the next orders we find similar situation.
New terms with the derivative $\pa_\th$ do not increase powers of $L^{-1}$.
As a result, we obtain solution of the form
$$
\psi_{l}\sim L^{1/2}\f{F_{(n)}}{K^n}\psi_0.
$$

Above we assumed that $C(\th )=\sum P_n (\cos^2 \th )$.
Because the action of $\pa_\th$ on $P_n (\cos^2\th )$ produces factor n,
in the higher orders in $O(C)/K$ in the terms $\Box_k \psi_n$,
in principle, accumulate  powers of $n$.  Trying to sum all orders, 
we encounter the problem of convergence.
To avoid this problem, we consider non-perturbative solution of  the equation
(\ref{d.2}).

We present $\Box$ as the sum $\Box_0 +\h\Box$, where $\Box_0$ is the 
zero-order part of d'Alambertian and $\h\Box$ written in notations 
(\ref{e.10})-(\ref{e.14}) is
\ba
\l{d.11}
&{}&\h\Box =
\f{1}{\sqrt{-\h{g}}}\pa_r \sqrt{-\h{g}} L\pa_r
- \f{1}{\sqrt{-g}}\pa_\th
  \sqrt{-g}\f{L^{1/2}b}{r\,f_1}\pa_r
- \f{1}{\sqrt{-g}}\pa_r \sqrt{-g}
  \f{L^{1/2}b}{r\,f_1}\pa_\th\\\nonumber
&+ &r^{-2}(f_1^2 -1 )(\pa_\th^2 +\cot\th\pa_\th ) +
\f{1}{\sqrt{-\h{g}} }\pa_\th \sqrt{-\h{g}}  (r\, f_1 )^{-2} \pa_\th +
r^{-2} (f_2^{-2}-1) \pa^2_\p
.\ea

 We look for a solution $\psi$  in the form $\psi_0 +\h\psi$,
where $\psi_0$ is the zero-order $s$-mode. Equation $\Box\psi =0$ takes the
form
\be
\l{k.1}
\Box_0\h\psi +\h\Box\psi_0 +\h\Box\h\psi =0
\ee
We take an ansatz for $\h\psi$ in a form $L^{1/2}\p\psi_0$, 
where $\p$ is sufficiently smooth function and 
$L^{1/2}\pa_r\p$ is finite in the limit $L=0$. 
We have
\be
\l{k.3}
\pa_r\h\psi =\pa_r( L^{1/2}\p\psi_0 ) =bibitem
\left(\f{M}{r^2}\pm i\o\right) L^{-1/2}\p\psi_0 +
 L^{1/2}(\pa_r\p)\psi_0 .
\ee
Let us consider  $\h\Box\psi_0$. The terms with derivatives $\pa_\th$ and $\pa_\p$
acting on $\psi_0$ yield zero. We have
\be
\l{k.4}
\h\Box\psi_0 =
\f{1}{\sqrt{-\h{g}}}L(\pa_r \sqrt{-\h{g}}) \pa_r\psi_0
+\f{1}{\sqrt{-g}}\pa_\th
  \sqrt{-g}\f{L^{1/2}b}{r\,f_1}\pa_r\psi_0 =
\pm i\o L^{-1/2}(F_1 + F_2 ) \psi_0,
\ee 
where  
$$
\f{1}{\sqrt{-\h{g}}}L(\pa_r \sqrt{-\h{g}}) \pa_r\psi_0
 =\pm i\o L^{-1/2}F_1 \psi_0 
$$
and
$$
\f{1}{\sqrt{-g}}\pa_\th
  \sqrt{-g}\f{L^{1/2}b}{r\,f_1}\pa_r\psi_0 =\pm i\o L^{-1/2}F_2  \psi_0
$$
Next, let us consider the term $\h\Box\h\psi $.
The  $(rr)$ term is 
$$
L\f{(\pa_r\sqrt{-\h{g}})}{\sqrt{-\h{g}}}\pa_r\h\psi  = 
 F_1\left(\f{M}{r^2} \pm i\o\right) L^{-1/2}\p\psi_0 +F_1 L^{1/2}(\pa_r \p )\psi_0
.$$
The $(\th r)$ term is 
$$\f{1}{\sqrt{-g}}\pa_\th         
  \sqrt{-g}\f{L^{1/2}b}{r\,f_1}\pa_r\h\psi
=L^{1/2}[(\pa_\th F_2) \pa_r \h\psi + F_2 \pa_r\pa_\th \h\psi ]=
L^{-1/2}\left(\f{M}{r^2} \pm i\o\right)[(\pa_\th F_2 )\p +F_2 (\pa_\th\p )]\psi_0
,$$
and the $(r\th)$ term is 
$$ \f{1}{\sqrt{-g}}\pa_r
  \sqrt{-g}\f{L^{1/2}b}{r\,f_1}\pa_\th\h\psi
= L^{-1/2}F_3\pa_\th\h\psi +L^{1/2}F_2\pa_r\pa_\th\h\psi 
.$$
Collecting the leading in $L^{-1}$ terms, we 
obtain Eq.(\ref{k.1}) as
\be
\l{k.2}
\Box_0\h\psi + L^{-1/2}(F_1 +F_2 )i\o\psi_0 
+ L^{-1/2}(\Psi_1 \p\psi_0 +\Psi_2\pa_\th \p\psi_0 )=0.
\ee
Eq.(\ref{k.2}) has the same functional form as Eq.(\ref{d.9}), and we conclude that
the ansatz $\h\psi =L^{1/2}\p\psi_0$ yields a solution of (\ref{k.2}).

\section{Isotropic  geodesics }

In this section we consider a class of the isotropic  geodesics 
in the metric (\ref{e.6}) depending on $r$ and $\th$
and show that in the vicinity of horizon they differ from geodesics in the 
Schwarzschild background by the terms of order $L^{1/2}= (1-2M/r)^{1/2}$.
Following the standard treatment \cite{chand} 
we start from the Lagrangian corresponding
to the metric (\ref{e.6}) which we present as 
\ba
\l{1.g}
2{\cal{L}}=-L{\dot{t}}^2 +\f{ {\dot{r}}^2 }{L}
\h{g}_{rr} 
+2\f{r\dot{r}\dot{\th}}{L^{1/2}}
\h{g}_{r\th } 
+r^2 \left[{\dot\th}^2 \h{g}_{\th\th}  +
 {\dot\p}^2 \sin^2\th   \h{g}_{\p\p}  
 \right]
.\ea
Derivatives are taken with respect to an affine parameter on geodesic.
The Lagrange equations following from the Lagrangian yield 
 the geodesic equations for $r(\tau ), \,t(\tau )$ 
and $\th(\tau ),\,\,\p(\tau )$. The equations for  $r(\tau ) $ and 
$\th(\tau )$ are
\ba
\l{2.g}
&{}&\f{\ddot{r}}{L} \h{g}_{rr} -\f{  {\dot{r}}^2 }{r^2 L^2 }\h{g}_{rr} 
+ \f{E^2 }{r^2 L^2}+ \f{  {\dot{r}}^2 }{2L}\pa_r \h{g}_{rr}+
\f{ \dot{r}\dot\th }{L}\pa_\th \h{g}_{rr}  +
\\\nonumber
&{}&+ \f{ \ddot{\th} r }
{L^{1/2}}\h{g}_{r\th} +\dot{\th}^2 \left(\f{r}{L^{1/2}}\pa_\th\h{g}_{r\th} 
-{r}\h{g}_{\th\th}-\f{r^2}{2}\pa_r\h{g}_{\th\th}\right)
-\dot{\p}^2 \left(r\h{g}_{\p\p} +\f{r^2}{2}\pa_r\h{g}_{\p\p} \right)
=0,
\\
\l{21.g}
&{}&\ddot\th r^2 \h{g}_{\th\th}+2r\dot\th\dot{r} \h{g}_{\th\th}+
r^2\dot\th\dot{r}\pa_r \h{g}_{\th\th} + 
\f{1}{2}r^2\dot\th^2\pa_\th \h{g}_{\th\th}+
\f{\ddot{r}r +\dot{r}^2}{L^{1/2}}\h{g}_{r\th}
-\f{\dot{r}^2}{rL^{3/2}}\h{g}_{r\th} +
\\\nonumber
&{}& +\f{\dot{r}^2 r}{L^{1/2}}\pa_r \h{g}_{r\th}-
\f{\dot{r}^2}{2L}\pa_\th \h{g}_{rr} -
\dot{\p}^2 r^2 (\sin\th\cos\th\h{g}_{\p\p} +
\sin^2\th \pa_\th \h{g}_{\p\p} ) =0.
\ea
Here $\pa_r \h{g}_{ij} = K\pa_K \h{g}_{ij} /L^{1/2}r$.
Because $t$ and $\p$ are cyclic variables, we solved the corresponding
equations and set in the Lagrange equations $\dot{t}= E/L$ and 
$r^2\dot\p\sin^2\th =B=const$. We consider the geodesics with  $B=0$.

For the isotropic geodesics equation ${\cal{L}}=0$ 
 is the first integral of the system of geodesic equations
\be
\l{3.g}
-\f{E^2}{2L}+\f{ {\dot{r}}^2 }{2L}\h{g}_{rr} +\f{r\dot{r}\dot{\th}}{L^{1/2}}
\h{g}_{r\th } +\f{r^2}{2} \dot\th^2 \h{g}_{\th\th} =0
.\ee

In its general form the system of equations is intractable. To proceed, we 
consider the near-horizon regon $r\rightarrow 2M$. 
In this limit $L\rightarrow 0$.
Examining the equations (\ref{2.g}), (\ref{21.g}) and (\ref{3.g}) we see that 
in the near-horizon limit we can look for a solution in the form
\ba
\l{81.g}
&{}&\dot{r}= C +C_1 L^{1/2}+\cdots,\\\nonumber
&{}&\dot\th =\f{A}{L^{1/2}}+A_1 +\cdots,\qquad 
\ddot{\th}=-\f{A\dot{r}}{r^2 L^{3/2}}+\cdots.
\ea
With the ansatz (\ref{81.g}) the leading in $L^{-1}$ 
terms in (\ref{2.g}) and (\ref{21.g}) are
\ba
\l{4.g}
\f{E^2}{r^2 L^2} -\f{\dot{r}^2}{r^2 L^2}\h{g}_{rr}
 +\f{\ddot{\th}r}{L^{1/2}}\h{g}_{r\th}=0,\\
\l{41.g}
\ddot{\th}r^2\h{g}_{\th\th}-\f{\dot{r}^2}{r L^{1/2}}\h{g}_{r\th}=0.
\ea
Substituting the ansatz (\ref{81.g}), we obtain
\ba
\l{9.g}
L^{-2}\, [E^2 -C^2 \bar{g}_{rr} -rAC \bar{g}_{r\th}] =0,\\
\l{91.g}
L^{-3/2}\,[C\bar{g}_{r\th} +rA\bar{g}_{\th\th}]=0.
\ea
and (\ref{3.g}) is
\be
\l{5.g}
L^{-1}[C^2 \bar{g}_{rr} -E^2 +2r CA \bar{g}_{r\th }
 +r^2  A^2\bar{g}_{\th\th}]=0.
\ee
Here we have introduced
$
\bar{g}_{ij}= \h{g}_{ij}|_{r=2M}$.  Note that  $\h{g}_{ij}$ 
are the components of the metric without the factors $L$ (see (\ref{e.11}).
Using the relation
\be
\l{10.g}
\h{g}_{rr} \h{g}_{\th\th} -\h{g}_{r\th}^2 =\h{g}_{\th\th}
,\ee
from the Eqs.(\ref{4.g}) and (\ref{41.g}) we obtain
\ba
\l{6.g}
C^2 =E^2,\\
\l{61.g}
A \simeq -\f{ C \bar{g}_{r\th } }{2M\bar{g}_{\th\th }}.
\ea
Because of the relation (\ref{10.g})
Eq. (\ref{5.g}) turns to identity.

 In the limit $b =0$  the Lagrangian
reduces to that of the Schwarzschild metric. 
In the spherically-symmetric metrics trajectories of the geodesics
are located in a plane going through
the symmetry center. 
Position of the plane depends on the initial conditions.
$\th$ and $\p$ are
coordinates in a coordinate system with the origin located at the symmetry
center.
Position of the plane  depends on the initial conditions.
Solution of the radial geodesic equations with the initial conditions
$\th (\tau_0 )=\pi/2,\quad\dot{\th} (\tau_0 )=0, 
\quad \p (\tau_0 =\dot\p (\tau_0 )=0 $ is
\ba
\l{7.g}
\dot{t}^2  = E^2 = \dot{r}^2\\\nonumber
\th (\tau ) = \pi/2,\\\nonumber
\p(\tau )  = 0.
\ea
Comparing solution (\ref{6.g}), (\ref{61.g}) with that in the Schwarzschild
metric and adjusting the integration constants, we obtain
\ba
\l{8.g}
&{}&r =2M +E\tau,\\\nonumber
&{}&t =t_0 + E\tau,\\\nonumber
&{}&\th \simeq\f{\pi}{2}-
\f{ \bar{g}_{r\th } }{2\bar{g}_{\th\th }}
\sqrt{\f{\tau E}{2M}}=\f{\pi}{2}-
\f{ \bar{g}_{r\th } }{2\bar{g}_{\th\th }}L^{1/2}.
\ea
In the vicinity of horizon solution (\ref{8.g}) differ from (\ref{7.g})
by the terms of order $L^{1/2}$.

\section{Conclusions}

A black hole
emerging as a result of the collaps  can be  
considered as practically  stationary,
the surface of the collapsing body approaches the horizon as $r-2M \sim$
$const e^{-t/2M}$ with a very small characteristic time ($\sim 2GM/c^3$ in
dimensionful units \cite{LL} ), and 
 propagation of the wave packets can be treated  as propagation in the 
background of the stationary black hole.
Evolution of the massless field is determined by the data at the past null
infinity $I^-$ in the basis $\{ u_i^{(-)}\}$.
Alternatively the field $\p$ can be expanded at
the hypersurface $\S^+ =I^+ \oplus H^+$ where $I^+$ is the future
null infinity and $H^+$ is the event horizon
\be
\l{2.2}
\p = \sum_i (b_i u^{(+)}_i + b^+_i u^{(+)*}_i +c_i q_i + c^+_i q^*_i  ),
,\ee
where  $\{u^{(+)}_i\}$ is the orthonormal set of modes which
contain at the $I^+$ only positive frequencies and $\{q_i\}$ is
the orthonormal set of solutions of the wave equation which
contains no outgoing components \cite{haw}.
The modes $\{u^{(+)}_i\}$ traced back to $I^-$ can be
expanded in terms of the modes $\{u^{(-)}_i\}$
$$
u_\o^{(+)}  = \int  d\o' (\a_{\o\o'}u_{\o'} ^{(-)} +
\b_{\o\o'}u_{\o'} ^{(-) *}).
$$
The Bogolubov coefficients $\b_{\o\o'}$ calculated at the surface $I_-$ are
\be
\l{h.2}
\b_{\o\o'} = i\int_{I_-} 4\pi r^2 dv u_\o^{(+)} (u(v))\stackrel
{\leftrightarrow}{\pa_v } u_{\o'}^{(-)} (v).
\ee
Actual calculation of $\b_{\o\o'}$ is performed in the limit $r\rightarrow 2M$.
 Because 
in the near-horizon region the geodesics have the same form as
in the Schwarzshild background the modes $u ^{(+)}$ traced back to $I^-$ 
are piled
in the near-horizon region  the
same as in the Schwarzschild case. At the horizon, the additional parts in
the modes depending on the supertranslation field are proportional to
$L^{1/2} =(1-2M/r )^{1/2}$ and  vanish. This is an accuracy of conventional
calculations of the Hawking effect (cf. \cite{brout}).
Thus, the Bogolubov coefficients and the density matrix of thermal radiation 
are the same as
in the Schwarzschild background.

{\large\bf Acknoledgments}

I am grateful to L. Slad and M. Smolyakov for useful discussions.

This work was partially supported by the Ministry of Science and Education of
Russian Federation under project 01201255504.


\begin{thebibliography}{99}

\bibitem{bondy}
H.Bondy, M.G.J.van der Burg, A.W.K.Metzner, {\it Gravitational vaves
in general relativity 7}, Proc. Roy. Soc. Lond. {\bf A269} 21 (1962).
\bibitem{sachs}
R.K.Sachs, {\it Gravitational vaves
in general relativity 8. Waves in asymptotically flat space-time.}
 Proc. Roy. Soc. Lond. {\bf A270} 103 (1962).
\bibitem{bar1}
G. Barnich and G. Compere, {\it Classical central extension for
asymptotic symmetries at null infinity in three spacetime dimensions},
Class. Quant. Grav. {\bf 24} F15, (2007) arXiv:gr-qc/0610130.
\bibitem{bar3}
G.Barnich and and C.Troessaert, {\it Aspects of the BMS/CFT correspondence},
JHEP 05 (2010) 062, arXiv:1001.1541.
\bibitem{bar2}
G.Barnich and and C.Troessaert, {\it Symmetries of asymptotically flat
4 dimensional spacetimes at null infinity revisited},
Phys. Rev. Lett. {\bf B105} 111103 (2010), arXiv:gr-qc/0909.2617.
\bibitem{strom1}
A. Strominger, {\it On BMS Invariance of Gravitational Scattering},
JHEP {\bf 1407} 152 (2014), arXiv:1312.2229.

\bibitem{comp2}
G.Compere and J.Long, {\it Classical static final state of collapse with
supertranslation memory},
Class. Quant. Grav. {\bf 33} (2016) 195001, arXiv:1602.05197.

\bibitem{strom2}
A. Strominger and A. Zhiboedov, {\it
Gravitational Memory, BMS Supertranslations and Soft Theorems},
JHEP 01, 086 (2016), arXiv:1411.5745.

\bibitem{winb}
S. Weinberg, {\it Infrared photons and gravitons}, Phys. Rev. {\bf 140}, B516,
(1965).

\bibitem{hawk}
S.W. Hawking, M.J. Perry, A. Strominger,
{\it Soft Hair on Black Holes}, 	
Phys. Rev. Lett. 116, 231301 (2016), arXiv:1601.00921.

\bibitem{comp1}
G.Compere and J.Long, {\it Vacua of the gravitational field}, JHEP 07 (2016) 137,
arXiv:1601.04958.

\bibitem{strom3}
A. Strominger, {\it Black Hole Information Revisited}, arXiv:1706.07143

\bibitem{haw}
S.W. Hawking, {\it Particle creation by black holes},
Comm. Math. Phys. {\bf 43} (1975) 199.

\bibitem{bir}
N.D. Birrell and P.C.W. Davies, {\it Quantum Fields in Curved Space},
Cambridge, England, 1982.

\bibitem{brout}
R. Brout, S. Massar, R. Parentani, P. Spindel, {\it A Primer for Black Hole
Quantum Physics},
Phys. Rept. {\bf 260} (1995) 329,  gr-qc/0710.4345

\bibitem{flet}
S.J. Fletcher and A.W.C. Lun, {\it The Kerr spacetime in generalized
Bondi-Sachs coordinates}, Class. and Quant. Grav. {\bf 20} (2003) 4153.


\bibitem{bar4}
G.Barnich and and C.Troessaert,	{\it BMS charge algebra},
JHEP 12 (2011) 003, arXiv:1309.0794.

\bibitem{sanch}
N.Sanchez
{\it Absorption and emission spectra of a Schwarzschild black hole}
Phys. Rev. {\bf D 18}, 1030 (1978).

\bibitem{viss}
M. Visser, {\it Essential and inessential features of Hawking radiation}
 	Int.J.Mod.Phys. {\bf D12} ,649, (2003), arXiv:hep-th/0106111.

\bibitem{chand}
S. Chandrasekhar, {\it The Mathematical Theory of Black Holes}, 1983,
Oxford University Press, New York.

\bibitem{LL}
L.D. Landau and E.M. Lifshitz, {\it The Classical Theory of Fields}, Moscow, 1988.


\end{thebibliography}
\end{document}